\documentclass[journal]{IEEEtran}
\IEEEoverridecommandlockouts

\usepackage[utf8]{inputenc}
\usepackage[english]{babel}
\usepackage[autostyle]{csquotes} 
\usepackage{url}

\usepackage{hyperref}
\usepackage{cuted}  
\usepackage{caption}
\usepackage{lipsum}
\usepackage{makecell}
\usepackage{multirow}
\usepackage{multicol}
\usepackage{cite}
\usepackage{amsmath,amssymb,amsfonts}
\usepackage{graphicx}
\usepackage{enumitem}
\usepackage{textcomp}
\usepackage{xcolor}
\usepackage{natbib}
\usepackage{array}
\usepackage{multirow}
\usepackage{booktabs} 
\usepackage{tabularx}
\usepackage{float}
\usepackage{colortbl}  
\usepackage{algorithm} 
\usepackage{algpseudocode}
\usepackage{amsmath}
\definecolor{lightgray}{gray}{0.9}
\newcolumntype{L}{>{\raggedright\arraybackslash}X}
\usepackage{algorithm}
\usepackage{algpseudocode}
\usepackage{booktabs} 
\usepackage{bbding}
\usepackage{tikz}
\usepackage{xcolor}

\usepackage{authblk}
\definecolor{myGreen}{HTML}{659E7D}
\definecolor{myOrange}{HTML}{D5653D}
\definecolor{myBlue}{HTML}{5E829A}

\tikzset{
    numcirc/.style={
        circle,
        inner sep=0pt,      
        minimum size=1em, 
        text=white,
        font=\bfseries\scriptsize, 
        anchor=base,
        yshift=0.5pt
    }
}

\newcommand{\cgreen}[1]{\tikz[baseline=(char.base)]{\node[numcirc, fill=myGreen] (char) {#1};}}
\newcommand{\corange}[1]{\tikz[baseline=(char.base)]{\node[numcirc, fill=myOrange] (char) {#1};}}
\newcommand{\cblue}[1]{\tikz[baseline=(char.base)]{\node[numcirc, fill=myBlue] (char) {#1};}}

\def\BibTeX{{\rm B\kern-.05em{\sc i\kern-.025em b}\kern-.08em
    T\kern-.1667em\lower.7ex\hbox{E}\kern-.125emX}}

\newcommand{\approach}{AgentGuard}
\newcommand{\fref}[1]{Fig.~\ref{#1}}
\newcommand{\tref}[1]{Table~\ref{#1}}
\newcommand{\sref}[1]{Section~\ref{#1}}

\begin{document}

\title{\approach: A Multi-Agent Framework for Robust Package Confusion Detection via Hybrid Search and Metadata-Content Fusion}

\author[1]{Yu Li}
\author[2]{Wei Ma}
\author[2]{Zhi Chen}
\author[2]{Ye Liu}
\author[2]{Lingxiao Jiang}
\author[3]{Junyi Tao}
\author[1]{Hao Liu}
\author[1]{Yongqiang Lyu}
\author[1]{Qiang~Hu\textsuperscript{\Envelope}}

\affil[1]{Tianjin University, China}
\affil[2]{Singapore Management University, Singapore}
\affil[3]{Amazon, United States}

\maketitle
\renewcommand{\thefootnote}{\Envelope}
\footnotetext{Corresponding author:\texttt{qianghu@tju.edu.cn}}
\renewcommand{\thefootnote}{\arabic{footnote}}
\begin{abstract}

The proliferation of open-source software (OSS) has made software supply chains prime targets for attacks like Package Confusion, where adversaries publish malicious packages with names deceptively similar to legitimate ones. To protect against such attacks and safeguard the use of OSS, multiple confusion detection methods have been proposed. However, existing methods are limited to single-signal retrieval strategies (relying solely on lexical or semantic metrics), struggle with high false positive rates (FPR), and are vulnerable to adversarial evasion. Critically, as content-agnostic approaches, they fundamentally fail to distinguish benign packages with high naming similarity from malicious, code-dissimilar impersonations, leading to persistent high FPR.

To address these limitations, we introduce AgentGuard, a novel multi-agents based framework for package confusion detection. Specifically, it first discovers potential confusion targets using fine-tuned word embedding models with hybrid similarity search. After that, It subsequently evaluates risk via a fused machine learning model that uniquely combines: (1) a multi-dimensional metadata group and (2) a novel package content analysis group, to reduce the FPR and mitigate the impact of adversarial evasion.

To assess the effectiveness of AgentGuard, we evaluate it on challenging ConfuDB and NeupaneDB datasets. Our results demonstrate that AgentGuard significantly outperforms state-of-the-art baselines, ConfuGuard and Typomind, improving precision by 12\%-49\% while simultaneously reducing the FPR by 11\%-35\%, and effectively discovers the confused package. 

\end{abstract}


\section{Introduction}
\label{sec:introduction}
Modern software development relies heavily on open-source package ecosystems, with registries like NPM~\cite{npm} and PyPI~\cite{pypi1} hosting millions of packages and serving billions of weekly downloads~\cite{npm2,spr,spr1}. This scale and openness create significant attack surfaces for software supply chain threats~\cite{supplychain1,supplychain3,supplychain2,supplychain4,supplychain5}. Among these, package confusion attacks pose a particularly insidious risk: adversaries publish malicious packages with names designed to closely resemble legitimate ones, deceiving developers into installing them and achieving arbitrary code execution. These attacks exploit diverse confusion mechanisms spanning lexical (typosquatting like \texttt{bz2file} and \texttt{bz2fiel}), syntactic (delimiter modifications, reordering or scope default like \texttt{bz2file} and \texttt{@xx/bz2file}), and semantic (synonym substitution like \texttt{bz2file} and \texttt{bzip}) levels~\cite{Typosquatting/confusion/Vu, neupane}.

To defend against these attacks and ensure the safe usage of OSS, multiple package confusion detection methods have been proposed. However, these methods have 
evolved from simple lexical matching to semantic analysis and metadata-based filtering, and therefore, are still facing critical limitations:

\paragraph{Ineffective Target Discovery} Existing methods typically rely on a single signal, such as purely semantic or purely lexical search, and thus, fail to capture the diversity of confusion attacks. For example, semantic-only search misses obvious lexical typos, while syntactic-only search is blind to semantic attacks, resulting in poor target recall.

\paragraph{High False-Positive Rates} Naming similarity is widespread. Crucially, existing content-agnostic detectors fundamentally struggle to distinguish malicious impersonation from legitimate packages that incidentally similar names and metadata, leading to excessive false positives and alert fatigue.

\paragraph{Susceptibility to Adversarial Evasion} Defenses relying on static rules or easily manipulable metadata features remain fragile against attackers who can forge or obfuscate these signals. To address these challenges, we introduce AgentGuard, a multi-agent framework that integrates a novel, fused classification model for package confusion detection. First, addressing the ineffective target discovery challenge, AgentGuard employs a hybrid (semantic and syntactic) search strategy. This strategy combines fine-tuned semantic embeddings for conceptual recall with syntactic trigram similarity for lexical precision, enabling it to discover confusion targets effectively. Second, focusing on the limitations of high FPR and adversarial evasion, we design a fused machine learning model that moves beyond metadata-only analysis. This classifier is fortified with two distinct, complementary feature sets: (1) a multi-dimensional metadata feature group, which incorporates robust temporal signals (TS) to resist forgery, and (2) our proposed package content (PC) feature group. The PC features provide a compelling signal by comparing package size differences, file list similarity, dependency graphs, and code embeddings to distinguish benign packages with malicious mimics, simultaneously boosting recall and precision.

To assess the effectiveness of \approach{}, we conduct comprehensive experiments on two commonly used datasets, ConfuDB and NeupaneDB. The results demonstrated that, compared to the SOTA methods ConfuGuard and Typomind, \approach{} achieved up to 49\% higher precision scores with 35\% lower FPR. Our ablation studies showcase the importance of our agent design and feature selection.   

To summarize, the main contributions of this paper are: 
\begin{itemize}[leftmargin=*]
\item We propose a hybrid search strategy that integrates semantic and syntactic signals to effectively discover confusion targets, outperforming single similarity metric approaches.
\item We introduce a novel package content features set~(comprising 4 features: packages size ratio, file list similarity, dependency similarity, and code embeddings similarity) that quantifies code-level plausibility to systematically reduce false positives and accurately identify malicious impersonations. 
\item We design a robust detection model grounded in multi-dimensional feature analysis. By combining the proposed package content signals with temporal and metadata quality features, our model establishes a resilient verification mechanism that effectively withstands adversarial metadata manipulation and forgery.
\item We design \approach{}, the first multi-agent based package confusion detection framework with SOTA performance and broad ecosystem adaptability. 
\end{itemize}

This article extends our preliminary work~\cite{agentguard}. Compared to the conference version, we introduce two novel modules to enhance \approach{}: 
First, to address the limitation of ineffective target discovery, we implement a \textit{hybrid search strategy} that autonomously and accurately identifies the correct legitimate target. 
Second, to fundamentally resolve the High FPR challenge, we introduce a \textit{deep package content analysis} module. This module performs content-aware verification by comparing package structure (e.g., size, file lists, dependencies) and code-based vector similarity against the discovered target. This critical enhancement enables the model to precisely distinguish malicious mimics from benign packages with high naming similarity, ensuring robust detection without excessive false alarms. 
Finally, we extensively evaluate these modules through controlled ablation studies, demonstrating that our fused approach achieves both superior baseline performance and best-in-class adversarial robustness.

\section{Background and Related Work}
\label{sec:background}

\subsection{Software Supply Chain Security} Software supply chain attacks have become a central challenge in modern cybersecurity~\cite{supplychain1,supplychain2,supplychain5,supplychain7}. Unlike traditional attacks, these methods contaminate third-party components, such as open-source packages, relied upon during development~\cite{SupplyChain6}. The event-stream incident is a canonical example of how injecting malicious code into a popular dependency can propagate threats to thousands of downstream projects.

Package confusion is a specific form of supply chain attack that targets the human developer through social engineering~\cite{confusionattack}. The concept of \textit{namesquatting} is well-studied in other domains, such as DNS and mobile application markets~\cite{DNS}. Package confusion adapts this idea to software package registries, where adversaries exploit naming similarities to deceive developers into installing malicious code~\cite{spr,spr1}. Consistent with established taxonomy~\cite{Typosquatting/confusion/Vu, neupane}, these attacks can be conceptually grouped into three broader categories: Lexical Confusion (e.g., typosquatting like \texttt{colour-string} and \texttt{color-string}), Syntactic Confusion (e.g., structural changes like \texttt{python-nmap} and \texttt{nmap-python}), and Semantic Confusion (e.g., synonym substitution like \texttt{bz2file} and \texttt{bzip}).

\subsection{Package Confusion Detection Strategies} \paragraph{Lexical and Syntactic Similarity} Early work focused on typosquatting~\cite{Typosquatting,Typosquatting1}. Taylor \emph{et al.}~\cite{Typosquatting/confusion/taylor}(the basis for TypoGard) and Vu \emph{et al.}~\cite{Typosquatting/confusion/Vu} applied Levenshtein distance to identify minor textual differences between package names in PyPI. Tools like Microsoft's OSSGadget~\cite{OSSGadget} also use lexical permutations to find squatting candidates. While effective against simple typographical errors, these methods fail to capture non-lexical attacks and often suffer from high FPR.

\paragraph{Semantic Similarity} A significant advancement came from Neupane \emph{et al.}~\cite{neupane} (the basis for Typomind) who pioneered the use of word embeddings (FastText~\cite{fasttext,fasttext1,fasttext_tool}) to compare package names at an abstract semantic level. Their work systematically categorized 13 types of confusion, demonstrating that many attacks are semantic, not just lexical. However, relying solely on name semantics still struggles to distinguish malicious intent from benign, functionally related packages, leading to persistent FPR challenges~\cite{lasa1,lasa2}.

\paragraph{Metadata-Based Analysis} To address the high FPR of name-only methods, research turned to package metadata. Zimmermann \emph{et al.}~\cite{Zimmermann} first showed that metadata properties, like maintainer count, correlate with security risks in the npm ecosystem. Ohm \emph{et al.}~\cite{ohm/sok} further explored using package information, dependencies, and scripts to train ML classifiers.

The state-of-the-art in this area is Jiang \emph{et al.}'s ConfuGuard~\cite{confuguard}. ConfuGuard introduced a comprehensive benignity filter based on 15 expert-derived heuristics (e.g., Distinct Purpose and Well-known Maintainers) to deeply analyze package metadata. This approach proved highly effective, reducing the FPR of its baseline detector from 80\% to 28\%. While powerful, ConfuGuard's detection model still operates on pre-identified candidate pairs and remains content-agnostic, making it difficult to accurately differentiate benign high-similarity packages from malicious mimics.

\paragraph{Code Content and Anomaly Analysis} Beyond static metadata, recent research has investigated code content and evolutionary dynamics to identify threats. Sejfia \emph{et al.}~\cite{Sejfia_2022} proposed a technique to detect malicious npm packages by identifying discrepancies between the published package content and its source code repository, leveraging the observation that malicious payloads are often absent from the public repository. Similarly, Garrett \emph{et al.}~\cite{garrett} focused on detecting suspicious updates by analyzing code drift, author changes, and API usage deviations between versions. Furthermore, Wyss \emph{et al.}~\cite{downloads/wyss} highlighted the complexity of the ecosystem by quantifying hidden code clones, demonstrating that many packages are legitimate forks rather than attacks. These works underscore the necessity of incorporating content-aware signals to differentiate malicious impersonations from benign clones.

\paragraph{Adversarial Robustness in Detection} A critical limitation of most metadata-based detectors is their reliance on \textit{honest} metadata. Halder \emph{et al.} (MeMPtec)~\cite{DTM/halder} were the first to systematically address this vulnerability. They introduced a framework classifying metadata features as Easy-to-Manipulate (ETM) (e.g., descriptions) or Difficult-to-Manipulate (DTM) (e.g., package age) based on properties of Monotonicity and Restricted Control. Their experiments confirmed that models relying on ETM features are fragile and suffer catastrophic performance collapse under simulated attacks, whereas DTM-inclusive models are significantly more robust.

\section{Design of AgentGuard}
\label{sec:method}

This section presents the design of AgentGuard. We begin by formalizing the problem and establishing our threat model, then articulate our design goals, followed by a system overview, and finally detail each agent's design.

\subsection{Problem Formulation, Threat Model and Design Goals}
\label{subsec:problem-threat}

\textit{\textbf{Problem Statement:}}
Given a newly published package $p$ with name $n_p$ in ecosystem $\mathcal{E}$, our objective is to determine whether $p$ is a confusion attack impersonating a legitimate package. We formulate this as an autonomous target discovery and verification task: given only the input $p$, the system must (1) identify a ranked set of potential legitimate targets $\mathcal{T} = \{t_1, t_2, ..., t_k\}$ from the entire ecosystem and pinpoint the most probable target $t_{best}$, and (2) extract and synthesize a holistic feature set fusing metadata plausibility with deep package content signals to accurately identify malicious confusion attacks against $t_{best}$.

\textit{\textbf{Threat Model:}}
Consistent with established taxonomy in software supply chain security~\cite{Typosquatting/confusion/Vu, ohm/sok}, we consider an adversary aiming to deceive developers into installing malicious packages by exploiting naming similarities with popular legitimate libraries. The adversary’s primary attack vector is the fabrication of a package name $n_p$ that is visually or cognitively similar to a target $n_t$. We assume the adversary employs diverse confusion techniques, spanning lexical (e.g., typosquatting), syntactic (e.g., delimiter tampering), and semantic (e.g., synonym substitution) domains, to evade detection and maximize the likelihood of accidental installation.

To enhance the deception's plausibility and bypass automated detectors, the adversary complements the primary name-based attack with secondary forgery strategies, operating on two dimensions: \textit{(1) Metadata Mimicry:} The adversary manipulates mutable metadata (corresponding to our MQ features), such as copying descriptions, forging repository URLs, or faking version numbers, to manufacture an appearance of legitimacy. \textit{(2) Malicious Payload Injection:} The adversary's ultimate goal is to execute malicious logic. Consequently, the resulting package content often exhibits a functional and structural divergence from the complex legitimate library it impersonates (e.g., a lightweight credential stealer mimicking a heavy utility library), resulting in significant disparities in file structure and code semantics.

Based on this, our model assumes the adversary can successfully forge signals related to metadata plausibility. However, we posit that the adversary operates under specific resource and logic constraints that prevent the trivial forgery of robust signals: \textit{(1) Temporal and Ecosystem Constraints:} We assume specific historical signals are immutable or prohibitively expensive to forge, such as the package's creation date and long-term release history. \textit{(2) Content-Functionality Trade-off:} We assume the adversary cannot replicate the legitimate target's code content while simultaneously introducing malicious behavior without leaving traces. A malicious impostor’s code deviates from the legitimate target in code size, dependency structure, and semantic vector representation, enabling reliable detection.

\textit{\textbf{Design Goals:}}
\label{subsec:design-goals}
To address the challenges identified in \sref{sec:introduction} and operate effectively under our threat model, AgentGuard is designed to satisfy the following objectives:
\begin{enumerate}[leftmargin=*]
    \item[G1] \textbf{Accurate Target Discovery.} Autonomously discover the correct legitimate target $t$ for any given input package $p$ from the entire ecosystem. This objective addresses the critical failures of single-signal retrieval which are blind to either semantic or syntactic attacks by implementing a hybrid search strategy. 
    \item[G2] \textbf{Content-Aware Detection.} Fundamentally verify the consistency between package identity and its functionality to alleviate FPR by deep package content analysis. This involves comparing the package's size, file list, dependencies, and average vector similarity of package code against the discovered legitimate target.
    \item[G3] \textbf{Robust Multi-Dimensional Detection.} Effectively distinguish malicious confusion from benign similarities by integrating a multi-dimensional features model. This model achieves resilience against adversarial metadata forgery by synthesizing signals from four complementary feature groups, crucially incorporating temporal (like package age) which are inherently difficult to forge.

    \item[G4] \textbf{Graceful Degradation of Metadata Acquisition and Classification Model.} AgentGuard ensures resilience across both data acquisition and classification models. For metadata acquisition, \approach{} needs to check a local metadata database and automatically fall back to the external API if data is unavailable or incomplete, ensuring robustness in data retrieval. For the classification model, AgentGuard employs a graceful degradation strategy. When the code embedding model does not support the target package's ecosystem, the model automatically defaults to the Metadata-Full (SS+MQ+CD+TS in RQ4) detector. This mechanism ensures controlled failure and continuous availability.

\end{enumerate}

\subsection{System Overview}
\label{subsec:overview}

\begin{figure*}[t]

    \centering
    \includegraphics[width=\linewidth]{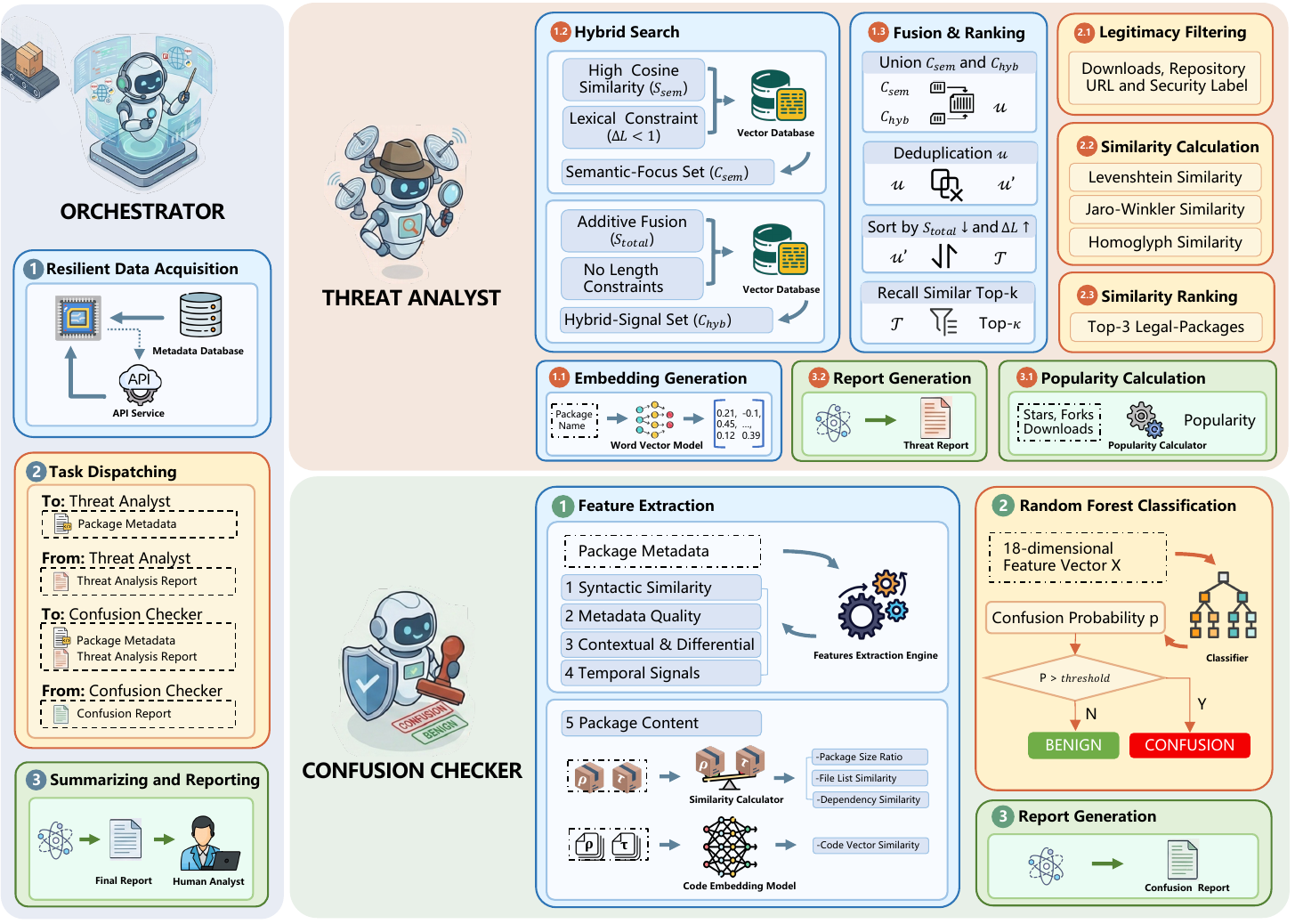}
    \caption{The overall architecture and workflow of the AgentGuard system. The Orchestrator Agent coordinates three specialized agents through a discovery-evaluation pipeline, processing packages from detection request to final threat assessment.}
    \label{fig:overview}

\end{figure*}

\textit{\textbf{Overview:}} AgentGuard is a multi-agent framework integrating LLM-based reasoning and tool-augmented ML for automated confusion detection. As shown in \fref{fig:overview}, it comprises three agents: the Orchestrator for coordination; the Threat Analyst for hybrid target discovery (\textbf{G1}); and the Confusion Checker for content-aware classification (\textbf{G2}, \textbf{G3}). A large-scale dataset (10M+ packages) underpins the system to ensure broad ecosystem adaptability (\textbf{G4}).

\textit{\textbf{Workflow:}} The workflow begins with the Orchestrator acquiring metadata for a target package~\cblue{1} and assigning it to the Threat Analyst~\cblue{2}. The Threat Analyst queries a vector database to retrieve the top-$k$ candidates~\corange{1.1}-\corange{1.3}, computes syntactic metrics (e.g., Levenshtein), and filters for legitimate targets based on popularity~\corange{2.1}-\corange{3.1} to output a structured threat report~\corange{3.2}. The Orchestrator then forwards this context to the Confusion Checker, which performs 18-feature extraction~\cgreen{1} and predicts confusion probability~\cgreen{2} to generate a diagnostic report~\cgreen{3}. Finally, the Orchestrator synthesizes all findings into a final report for human review~\cblue{3}. 

\textit{\textbf{Summary:}} By orchestrating specialized models including scalable embedding models for target discovery, deep code embeddings for semantic analysis, and lightweight classifiers for decision making, AgentGuard integrates analytical depth with operational robustness, while leveraging adaptive LLM-based reasoning for coordination.

\subsection{Orchestrator Agent}
\label{subsec:orchestrator}

The Orchestrator Agent serves as the central coordinator of AgentGuard, managing task initialization, inter-agent communication, and result aggregation. 
Powered by the LLM for adaptive decision-making, it bridges high-level reasoning with tool-assisted data operations.
Given an input package $p$ with name $n_p$ and ecosystem $\mathcal{E}$, it is in charge of the following tasks:

\noindent\textit{\textbf{Step 1:} Resilient Data Acquisition.} 
The Orchestrator first retrieves package metadata from our constructed database that includes 10,015,794 package metadata spanning 34 distinct ecosystems in \tref{tab:ecosystems} (\textbf{G4}). If local data is unavailable or incomplete, it automatically falls back to the API service provided by \texttt{Libraries.io}~\cite{libraries_io,ecosystems}, ensuring robustness in data retrieval.

    \begin{table}[tbp]
        \centering
        \caption{Metadata database covers package ecosystems and their package number.}
        \label{tab:ecosystems}
        
        \small 
    
        \begin{tabular*}{\columnwidth}{ 
            l @{\hspace{1em}} 
            r @{\hspace{2.5em} \@{\extracolsep{\fill}}} 
            | @{\hspace{2.5em} \@{\extracolsep{\fill}}} 
            l @{\hspace{1em}} 
            r 
        }
            
            \noalign{\hrule height 1pt} 
            \noalign{\smallskip}
            
            \textbf{Ecosystem} & \textbf{Number} & \textbf{Ecosystem} & \textbf{Number} \\
            \noalign{\smallskip}
            
            \hline 
            \noalign{\smallskip}

            actions & 31558 & hackage & 17810 \\
            adelie & 7572 & hex & 16192 \\
            alpine & 246740 & homebrew & 7462 \\
            bioconductor & 2379 & julia & 10957 \\
            bower & 70212 & maven & 501606 \\
            cargo & 151425 & npm & 4273272 \\
            carthage & 2062 & nuget & 639947 \\
            clojars & 20333 & packagist & 419316 \\
            cocoapods & 96394 & postmarketos & 2414 \\
            conda & 23555 & pub & 53548 \\
            cpan & 40282 & puppet & 7580 \\
            cran & 23563 & pypi & 576657 \\
            deno & 5431 & racket & 2508 \\
            docker & 1001906 & rubygems & 193051 \\
            elm & 2819 & spack & 8186 \\
            elpa & 677 & swiftpm & 8274 \\
            go & 1547656 & vcpkg & 2450 \\
            \noalign{\smallskip}
            
            \noalign{\hrule height 1pt} 
            
        \end{tabular*}
    \end{table}

\noindent\textit{\textbf{Step 2:} Task Dispatch and Conditional Scheduling.} 
The Orchestrator constructs a detection request $\langle n_p, \mathcal{E} \rangle$ and dispatches it to the Threat Analyst Agent for target discovery. Upon receiving the threat report (containing identified targets $\mathcal{T}$), it forwards the enriched task (including $p$, $\mathcal{T}$, and all metadata) to the Confusion Checker Agent for features extraction and classification for generating final confusion report.

\noindent\textit{\textbf{Step 3:} Result Aggregation and Reporting.} 
The Orchestrator consolidates the threat report and confusion report (if available) into a comprehensive assessment and forwarding it to the real-world human analyst for final decision-making.

\begin{table*}[ht!]
  \centering
  \caption{The 18 multi-dimensional features extracted by the Confusion Checker Agent, grouped into five categories.}
  \label{tab:features}

  \begin{tabularx}{\textwidth}{@{} l @{\hspace{1em}} l @{\hspace{1.5em}} L @{}}
    \toprule
    \textbf{Feature Group} & \textbf{Feature Name} & \textbf{Description} \\
    \midrule

    \multirow{3}[3]{*}{\shortstack[l]{\bfseries Syntactic Similarity (SS)}}
    & Maximum Levenshtein Similarity & Maximum Levenshtein similarity between the target and the Top-3 legitimate. \\
    \cmidrule(r){2-3} 
    & Maximum Jaro-Winkler Similarity & Maximum Jaro-Winkler similarity between the target and the Top-3 legitimate.\\
    \cmidrule(r){2-3} 
    & Maximum Homoglyph Score & Highest visual character similarity (e.g., 'o' vs '0') to Top-3 legitimate packages. \\
    
    \midrule 
    \multirow{5}[5]{*}{\shortstack[l]{\bfseries Metadata Quality (MQ)}}
    & Maintainer Adequacy & Signals if the package is actively managed and maintained, flag by 0 and 1.  \\
    \cmidrule(r){2-3} 
    & Repository URL Validity & Signals if repository URL is valid, non-empty, flag by 0 and 1.   \\
    \cmidrule(r){2-3} 
    & Version Format Validity & Signals if version string is legitimate and valid, flag by 0 and 1.  \\
    \cmidrule(r){2-3} 
    & License Validity & Signals if license is specified and legitimate, flag by 0 and 1. \\
    \cmidrule(r){2-3} 
    & Version Count & Total number of historical versions published for the package. \\
    
    \midrule

    \multirow{3}[3]{*}{\shortstack[l]{\bfseries Contextual \& Differential (CD)}}
    & High-Similarity Count & Number of candidates with syntactic similarity higher than threshold. \\
    \cmidrule(r){2-3} 
    & Minimum Length Difference Ratio & Smallest name length difference ratio relative to Top-3 candidates. \\
    \cmidrule(r){2-3} 
    & Target Package Popularity & Normalized popularity score [0, 1] of the target package itself. \\
    
    \midrule

    \multirow{3}[3]{*}{\shortstack[l]{\bfseries Temporal Signals (TS)}}
    & Package Age & Log-transformed age in days since package creation. \\
    \cmidrule(r){2-3} 
    & Time Since Last Release & Log-transformed days since the last version was published. \\
    \cmidrule(r){2-3} 
    & Time Since Last Update & Log-transformed days since the package metadata was last updated. \\
    
    \midrule
 
    \multirow{4}[4]{*}{\shortstack[l]{\bfseries Package Content (PC)}}
    & Package Size Ratio & Ratio of log-transformed code size of target package and legitimate package.\\
    \cmidrule(r){2-3} 
    & File List Similarity & Similarity of the file manifests of target package and legitimate package. \\
    \cmidrule(r){2-3} 
    & Dependency Similarity & Similarity of the dependency sets of target package and legitimate package. \\
    \cmidrule(r){2-3} 
    & Code-based Vector Similarity & Similarity of the mean code vectors of target package and legitimate package. \\
    
    \bottomrule
  \end{tabularx}
\end{table*}

\subsection{Threat Analyst Agent}
\label{subsec:threat-analyst}

The Threat Analyst Agent addresses autonomous target discovery, transforming single-input detection into a comparison problem by identifying potential legitimate targets $\mathcal{T}$ that a suspicious package may impersonate. 
Guided by the LLM, it coordinates semantic embedding queries and syntactic similarity analyses, balancing semantic breadth (capturing synonym-based confusion) with syntactic precision (detecting typosquatting) (\textbf{G1}).

\noindent\textit{\textbf{Step 1:} Hybrid Search for Candidate Generation.} The agent initiates the process by encoding the input package name $n_p$ using a fine-tuned word embedding model. This model leverages a subword-based mechanism to capture morphological variations, essential for handling diverse naming conventions across 34 ecosystems. To generate a precise set of potential targets, the agent employs a hybrid search strategy. Specifically, we calculate the \textit{Vector Cosine Similarity} between pre-computed target package name embeddings to capture semantic relatedness, and simultaneously compute \textit{Trigram Similarity} (based on the Dice coefficient of character trigrams) to quantify syntactic overlap and detect typosquatting~\cite{Typosquatting}. Leveraging these two metrics, the agent implements a dual-channel recall strategy to address the limitations of single-signal retrieval. As outlined in Algorithm 1, this workflow consists of two phases: Candidate Generation and Rank Fusion. In the Candidate Generation phase (Lines 1-2), we execute two parallel retrieval queries to maximize coverage. The first channel, the \textit{Semantic-Focus Set} ($\mathcal{C}_{sem}$), prioritizes packages with high vector cosine similarity ($S_{sem}$). Crucially, to filter out irrelevant semantic neighbors, we enforce a strict lexical constraint, retaining only candidates with a name length difference of at most $n$ characters ($\Delta L \le n$). The second channel, the \textit{Hybrid-Signal Set} ($\mathcal{C}_{hyb}$), retrieves candidates based on the additive fusion of semantic and trigram syntactic scores ($S_{syn}$), capturing targets with strong dual evidence without length constraints. In the Fusion and Ranking phase (Lines 3-8), we compute the union of these two sets. A critical deduplication step ensures that unique candidates are retained. Finally, the unified pool is re-ranked lexicographically: primarily by maximizing the combined similarity score ($S_{total}$) to prioritize high-confidence targets, and secondarily by minimizing the length difference ($\Delta L$) to favor lexically plausible impersonations (Line 9). The top-$k$ candidates form the final target set $\mathcal{T}$ (Line 10).

\algrenewcommand\algorithmicrequire{\textbf{Input:}}
\algrenewcommand\algorithmicensure{\textbf{Output:}}
\begin{algorithm}[t]
\caption{Hybrid Candidate Generation and Ranking}
\label{alg:hybrid_search}
\begin{algorithmic}[1]
\Require $p$: input package; $\mathcal{E}$: ecosystem;
\Statex \hskip1.5em $n$: max length diff threshold;
\Statex \hskip1.5em $K_1, K_2$: retrieval depth for semantic/hybrid channels;
\Statex \hskip1.5em $k$: final top-$k$ candidates output size;

\Ensure Ranked list of legitimate targets $\mathcal{T}$.

\vspace{0.3em}
\Statex \textit{\textbf{Phase 1: Dual-Channel Candidate Generation}}
\Statex \hskip0em $\triangleright$ \textit{Channel 1: Strict Semantic (High Precision)}
\State $\mathcal{C}_{sem} \gets \textsc{Retrieve}(\mathcal{E}, p, \text{score}\!=\!S_{sem}, \text{top}\!=\!K_1)$
\Statex \hskip2em \textbf{s.t.} $|\text{len}(p) - \text{len}(q)| \le n$ 

\Statex \hskip0em $\triangleright$ \textit{Channel 2: Broad Hybrid (High Recall, $K_2 > K_1$)}
\State $\mathcal{C}_{hyb} \gets \textsc{Retrieve}(\mathcal{E}, p, \text{score}\!=\!(S_{sem} + S_{syn}), \text{top}\!=\!K_2)$

\vspace{0.3em}
\Statex \textit{\textbf{Phase 2: Fusion and Ranking}}
\State $\mathcal{U} \gets \mathcal{C}_{sem} \cup \mathcal{C}_{hyb}$ 
\State \textbf{Deduplicate} $\mathcal{U}$ based on unique package names

\For{each candidate $q \in \mathcal{U}$}
    \State $S_{total}(q) \gets S_{sem}(p,q) + S_{syn}(p,q)$
    \State $\Delta L(q) \gets |\text{len}(p) - \text{len}(q)|$
\EndFor

\Statex \hskip0em $\triangleright$ \textit{Hierarchical Sort: Maximize Score, then Minimize Length Diff}
\State $\mathcal{T} \gets \textsc{Sort}(\mathcal{U})$ descending by tuple key: 
\Statex \hskip3em $\langle S_{total}(q), -\Delta L(q) \rangle$ 

\State \textbf{return} Top-$k$ elements of $\mathcal{T}$
\end{algorithmic}
\end{algorithm}

\noindent\textit{\textbf{Step 2:} Legitimate-Package Filtering and Ranking.}
Since attackers typically mimic popular packages to maximize impact, the agent filters candidates by download counts and repository verification. For each retained candidate $c_i$, it computes three syntactic similarity scores: Levenshtein distance~\cite{Levenshtein}, Jaro–Winkler similarity~\cite{Winkler}, and Homoglyph score~\cite{homo}. Candidates are ranked by $\max\{\text{Lev}, \text{Jaro}, \text{Homo}\}$, and the top three form the final legitimate package set ${T} = \{t_1, t_2, t_3\}$.

\noindent\textit{\textbf{Step 3:} Popularity-Driven Target Selection and Reporting.}
To pinpoint the single most probable victim $t_{best}$ from the candidate set ${T}$, the agent incorporates a popularity-weighted decision mechanism, based on the premise that attackers preferentially target high-value libraries to maximize impact.
First, the agent computes a normalized popularity score $S_{pop}(t_i) \in [0, 1]$ for each candidate $t_i \in {T}$ by aggregating download counts, stars, forks, and dependents~\cite{downloads/wyss}. 
Next, it re-ranks the candidates by synthesizing the syntactic similarity score (from Step 2) with this popularity score. The candidate with the highest combined score is identified as the $t_{best}$).
Finally, the agent packages $t_{best}$ along with its metadata and popularity profile into a \textit{threat report} for the Orchestrator.

\subsection{Confusion Checker Agent}
\label{subsec:confusion-checker}

The Confusion Checker Agent performs the final stage of analysis, moving beyond metadata to conduct a fused evaluation of confusion likelihood. This agent complements the LLM-driven reasoning of upstream components by employing a machine learning classifier for reliable classification. This classifier ensures reliability and interpretability by basing its decision on a holistic feature set, integrating not only metadata and similarity results, but also deep package content signals including packages size ratio, file structural similarity, dependency analysis, and semantic embeddings derived from the package source code (\textbf{G2}).

\noindent\textit{\textbf{Step 1:} Multi-Dimensional Feature Extraction.}
This feature extraction process is designed to overcome the core challenges of high FPR and adversarial evasion identified in \sref{sec:introduction}. 
Specifically, the agent extracts a comprehensive 18-dimensional feature vector $X\in\mathbb{R}^{18}$ organized into five feature groups in \tref{tab:features}. The features span Syntactic Similarity (SS) (name-level resemblance), Metadata Quality (MQ) (maintenance and validity signals), Contextual and Differential (CD) (comparative ecosystem context), Temporal Signals (TS) (like package age and creation duration that resist forgery), and critically, Package Content (PC) features group (analyzing package size ratio, file list similarity, dependency similarity, and codes similarity).
This fused design provides a holistic view, using temporal signal features for resilience against metadata forgery (\textbf{G3}) and package content features to definitively resolve the ambiguity in high-similarity scenarios and ensure precise classification(\textbf{G2}).

\noindent\textit{\textbf{Step 2:} Random Forest Classification.} The 18-dimensional feature vector $X \in \mathbb{R}^{18}$ is input into the final, deployed Random Forest (RF) classifier. This model's hyperparameters were optimized using the 5-fold stratified cross-validation. The final classifier was then trained on the entire combined dataset (all 5 folds) using these optimal settings. The RF model aggregates individual tree predictions via majority voting to output a single confusion probability, $\hat{P} = P(\text{Confusion} | X) \in [0, 1]$. This probability serves as both the input to the classification decision and as a confidence estimate for prioritizing human review.

\noindent\textit{\textbf{Step 3:} 
Threshold-Based Decision and Report Generation.}
The Confusion Checker Agent's final step is to translate the confusion probability $\hat{P}$ into a definitive classification. It compares the probability $\hat{P}$ from {\textit{Step 2} against this \textit{optimized threshold}. If $\hat{P}$ is greater than the threshold determined via F1-maximization in \sref{sec:exp}, the package is classified as \textit{Confusion}; otherwise, as \textit{Benign}. The agent generates a \textit{Confusion Report} with the classification decision and decision reason for manual review which is returned to the Orchestrator.

Through this modular multi-agent design, AgentGuard achieves a balance between scalability, analytical precision, and robustness against adversarial behavior. The empirical validation is presented in \sref{sec:exp}.

\section{EXPERIMENTS}
\label{sec:exp}
\subsection{Experimental setups}
Our all experiments were performed on a server equipped with a 32-core Intel Xeon Gold 6459C CPU, 64 GB of RAM, and 1 TB of storage. The training
and fine-tuning of word embedding model do not require GPUs.

\textit{\textbf{Baselines:}} We compared AgentGuard against two established baselines: \textit{Typomind}~\cite{neupane}, which utilizes semantic embeddings alongside lexical analysis to detect confusion primarily on package pairs, and \textit{ConfuGuard}~\cite{confuguard}, a state-of-the-art system that employs deep metadata analysis and a \enquote{Benignity Filter} to reduce false positives, often analyzing pre-identified candidate pairs.

\textit{\textbf{Datasets:}} Two primary datasets were used for evaluation. \textit{ConfuDB}~\cite{confuguard} comprises 2,361 packages triaged by security analysts during the development and deployment of related tools, representing more complex, real-world scenarios with a mix of confirmed attacks, stealthy threats, and benign packages. \textit{NeupaneDB}~\cite{neupane} includes 1,840 packages primarily derived from previous research, containing a larger proportion of confirmed, publicly documented confusion attack cases.

\textit{\textbf{Parameter Settings:}} For the autonomous target discovery (Algorithm \ref{alg:hybrid_search}), we empirically set the hyperparameters based on a preliminary grid search to balance recall and efficiency. We set the length difference threshold $n=1$ to strictly filter semantic neighbors. The retrieval depths are set to $K_1=100$ for the semantic-focus channel and $K_2=150$ for the hybrid-signal channel, acknowledging that the hybrid channel requires a broader search window to capture diverse confusion forms. The final candidate size is set to $k=10$.

\textit{\textbf{Word Embedding Model:}} To facilitate the Threat Analyst Agent's semantic discovery, we fine-tuned a FastText (cc.en.300.bin)~\cite{fasttext,fasttext1,fasttext_tool} model on a comprehensive corpus of 10,015,794 package names spanning 34 distinct ecosystems. This large-scale, cross-ecosystem tuning adapted the model to capture specific semantic nuances and subword structures prevalent in diverse package naming conventions.

\textit{\textbf{Code Embedding Model:}} The Confusion Checker Agent utilizes a pre-trained deep code embedding model, CodeBERT~\cite{codebert}, which acts as a powerful feature extractor for package content. This model was pre-trained on source code from six languages, including Python, JavaScript, Go, Java, PHP and Ruby. For both the input package and its discovered legitimate package, the agent generates embeddings for their respective source files. These file-level embeddings are then aggregated (via averaging) to create a single vector representing each package. The similarity between these two code-based vectors is then computed as the code similarity feature, providing a critical signal to the Random Forest classifier to identify whether the package is malicious confusion.

\textit{\textbf{Classification Model:}} AgentGuard's Confusion Checker Agent utilizes a Random Forest classifier~\cite{random}. Our model development and evaluation follow a rigorous two-stage process to ensure robust, unbiased performance metrics and an optimized final model for deployment: \textcircled{1}K-Fold Cross-Validation for Evaluation~\cite{Kfold,Kfold1}. To robustly evaluate the generalization performance of our 18-feature methodology and select optimal settings like the classification threshold, we employed 5-fold stratified cross-validation. This process generates a complete set of out-of-fold (OOF) predictions (where each sample is predicted by a model that did not see it during training)~\cite{oof,oof1}. All evaluation metrics reported in this paper including the performance in \tref{tab:baseline} and \tref{tab:ablation_study}, and the curves in \fref{fig:threshold} and \fref{fig:roc} are derived from these robust OOF predictions.
\textcircled{2}Final Model Training for Deployment. Following the successful validation in Stage 1, a single, final model was then trained using the selected optimal settings on the entire combined dataset. This final model is the artifact that is deployed within the live AgentGuard system for classifying new, previously unseen packages.

\subsection{Research Questions}
To comprehensively evaluate AgentGuard, we designed experiments addressing
three research questions:

\textit{\textbf{RQ1 (Overall Capability of \approach{}):}} How does AgentGuard select its classification threshold, and what is its overall classification capability?

\textit{\textbf{RQ2 (Performance Comparison with Baselines):}} How does AgentGuard compare to state-of-the-art tools (Typomind, ConfuGuard) in detecting real-world confusion attacks?

\textit{\textbf{RQ3 (Hybrid Search Strategy Capacity):}} How effective is AgentGuard's hybrid search target discovery at finding the correct legitimate target, compared to using purely semantic or purely syntactic search strategies?

\textit{\textbf{RQ4 (Feature Contribution and Robustness):}} What is the contribution of each feature category (SS, MQ, CD, TS and PC), and how robust is the model against adversarial metadata manipulation?

\subsection{RQ1 - Threshold Selection and Model Capability}
\textit{\textbf{Methodology:}}
To determine the optimal classification threshold, we employed a data-driven grid search over the [0.01, 0.99] range based on unbiased out-of-fold probabilities generated via 5-fold stratified cross-validation~\cite{Kfold,Kfold1}. We selected the threshold maximizing the F1-score independently for \textit{ConfuDB} and \textit{NeupaneDB} to accommodate their distinct data characteristics.

\textit{\textbf{Results:}} As shown in the relationship plots of F1-score with threshold (\fref{fig:threshold}), the F1-score for each dataset peaks at a different optimal threshold.
For \textit{ConfuDB}, as a more complex, real-world dataset, the F1-score peaks at an optimal threshold of 0.42. For \textit{NeupaneDB}, as a dataset of known attacks, the F1-score peaks at a threshold of 0.52.

This deviation from the conventional 0.5 threshold reflects the asymmetric cost of confusion detection.The fact that each dataset has its own optimal threshold validates our data-driven approach and highlights the different operating characteristics required for triaging real-world alerts (\textit{ConfuDB}) versus detecting known attack patterns (\textit{NeupaneDB}). For our final deployment model, we selected the threshold derived from \textit{ConfuDB} (0.42) as it is more representative of the practical requirement to minimize security risks in noisy, real-world monitoring.

\begin{figure}[tbp]
    \centering
    \includegraphics[width=\columnwidth]{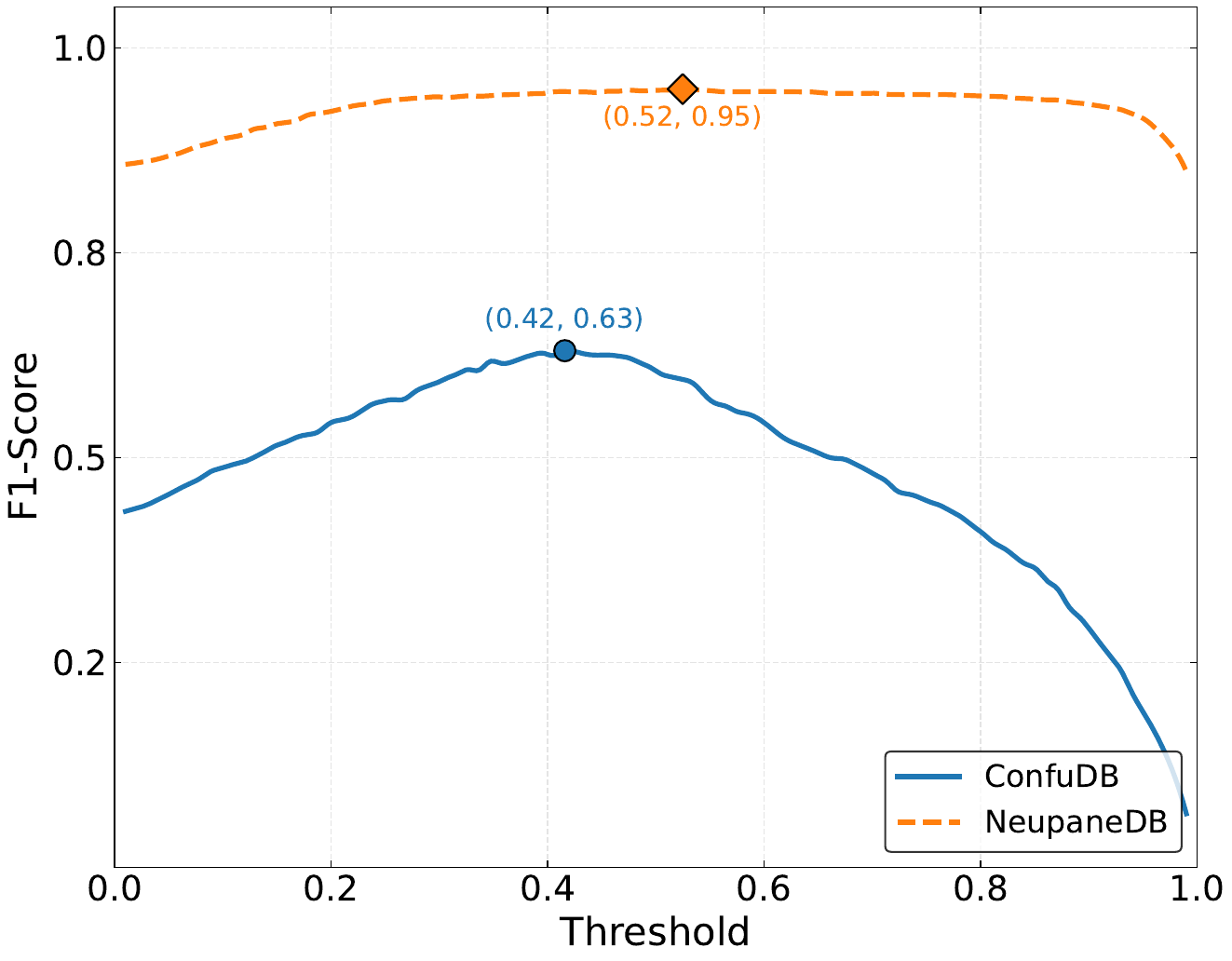}
    \caption{Relationship of F1-score with threshold. The plot identifies the best F1-Score (0.63 and 0.95) and its corresponding optimal threshold for each dataset.}
    \label{fig:threshold}
\end{figure}

\textit{\textbf{Overall Classification Capability:}} The ROC curves (\fref{fig:roc}), also plotted using the separate OOF predictions, demonstrate the model's robust classification capability in both contexts. A Receiver Operating Characteristic (ROC)~\cite{roc} curve plots the model's True Positive Rate (TPR) against its FPR across all possible thresholds. The Area Under the Curve (AUC)~\cite{auc} summarizes this plot into a single metric representing the probability that the model will correctly rank a random positive sample higher than a random negative one; an AUC of 1.0 signifies a perfect classifier, while 0.5 signifies random chance.

On ConfuDB, the model achieves a strong AUC of 0.83. On NeupaneDB, the model achieves an excellent AUC of 0.96. This significant difference in AUC scores confirms that ConfuDB represents a much more challenging classification problem due to the subtle differences between attacks and legitimate, similarly-named packages. The high AUC on both datasets verifies the model's strong discriminative performance.

\textit{\textbf{Answer to RQ1:}} \approach{} shows strong package confusion detection capability with AUC scores of 0.83 and 0.96 on ConfuDB and NeupaneDB, respectively.

\subsection{RQ2 - Performance Comparison with Baselines} \textit{\textbf{Overall Performance:}} We evaluated AgentGuard against two state-of-the-art baselines (Typomind and ConfuGuard) on both ConfuDB (2,337 packages, analyst-triaged real-world cases) and NeupaneDB (1,841 packages, publicly documented attacks). \tref{tab:baseline} presents the comprehensive results. AgentGuard consistently outperforms both baselines across all metrics on both datasets, achieving improvements of 12\%-49\% in precision, 16\%-27\% in recall/accuracy, and 17\%-43\% in F1-score (\textbf{G2}).

\begin{figure}[tbp]
    \centering
    \includegraphics[width=\columnwidth]{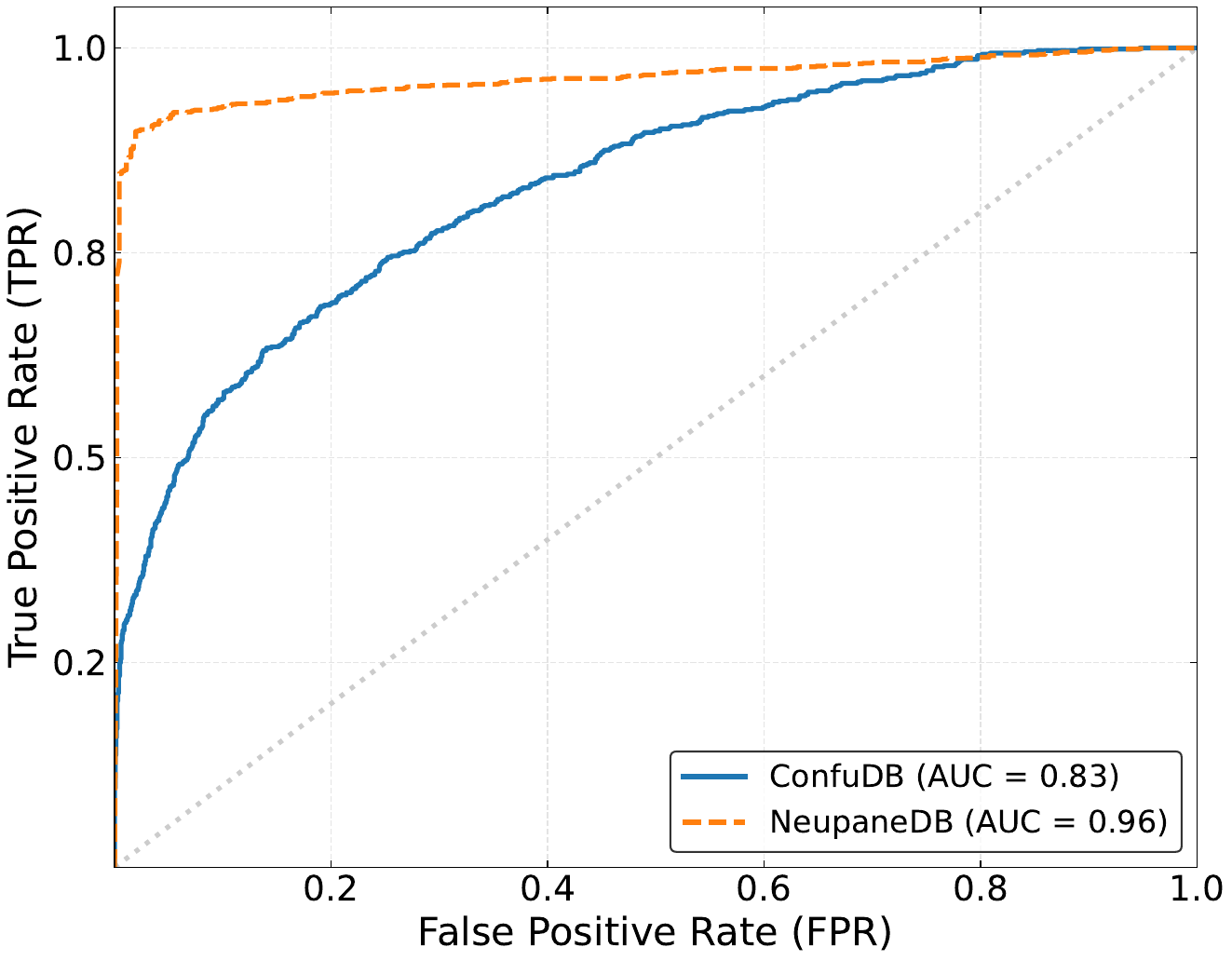}
    \caption{ROC curve showing FPR and TPR trade-off. The model achieves high AUC scores of 0.83 on ConfuDB and 0.96 on NeupaneDB.}
    \label{fig:roc}
\end{figure}

\begin{table*}[t]
\centering
\caption{Experiments result for AgentGuard with Typomind and ConfuGuard on the ConfuDB and NeupaneDB datasets}
\label{tab:baseline}
\renewcommand{\arraystretch}{1.1}

\resizebox{\textwidth}{!}{%
\begin{tabular}{ll cccc cccc cccc c} 
\toprule

\multirow{2}{*}{\textbf{Dataset}} & \multirow{2}{*}{\textbf{Class}} & \multicolumn{4}{c}{\textbf{AgentGuard}} & \multicolumn{4}{c}{\textbf{ConfuGuard}} & \multicolumn{4}{c}{\textbf{Typomind}} & \multirow{2}{*}{Support} \\ 
\cmidrule(lr){3-6} \cmidrule(lr){7-10} \cmidrule(lr){11-14}
& & Prec. & Recall & F1 & Acc. &  Prec. & Recall & F1 & Acc. & Prec. & Recall & F1 & Acc. & \\ 
\midrule 
\multirow{3}{*}{\textbf{ConfuDB}} & Benign & 0.86 & 0.86 & 0.86 & 0.86 & 0.40 & 0.51 & 0.45 & 0.51 & 0.31 & 0.69 & 0.30 & 0.69 & 1704 \\ 
\cmidrule(l){2-15} 
& Confusion & 0.63 & 0.64 & 0.63 & 0.64 & 0.55 & 0.53 & 0.49 & 0.53 & 0.31 & 0.31 & 0.52 & 0.31 & 633 \\ 
\cmidrule(l){2-15}

& Weight avg & \cellcolor{lightgray}\textbf{0.80} & \cellcolor{lightgray}\textbf{0.79} & \cellcolor{lightgray}\textbf{0.79} & \cellcolor{lightgray}\textbf{0.79} & 0.44 & 0.52 & 0.46 & 0.52 & 0.31 & 0.61 & 0.36 & 0.61 & 2337 \\ 
\midrule
\multirow{3}{*}{\textbf{NeupaneDB}} & Benign & 0.80 & 0.95 & 0.87 & 0.93 & 0.87 & 0.62 & 0.67 & 0.62 & 0.38 & 0.84 & 0.81 & 0.84 & 461 \\ 
\cmidrule(l){2-15}
& Confusion & 0.98 & 0.92 & 0.95 & 0.92 & 0.79 & 0.82 & 0.79 & 0.82 & 0.74 & 0.68 & 0.65 & 0.68 & 1380 \\ 
\cmidrule(l){2-15}

& Weight avg & \cellcolor{lightgray}\textbf{0.94} & \cellcolor{lightgray}\textbf{0.93} & \cellcolor{lightgray}\textbf{0.93} & \cellcolor{lightgray}\textbf{0.93} & 0.82 & 0.77 & 0.76 & 0.77 & 0.65 & 0.72 & 0.69 & 0.72 & 1841 \\ 
\bottomrule
\end{tabular}
} 
\end{table*}

\textit{\textbf{Dataset-Specific Analysis:}} On ConfuDB, the most challenging real-world dataset, AgentGuard achieves a weighted F1 of 0.79, representing a 72\% and 119\% relative improvement over ConfuGuard (0.46) and Typomind (0.36), respectively. This gap is driven by AgentGuard's superior ability to identify actual attacks. AgentGuard's Confusion Recall (0.64) substantially exceeds both ConfuGuard (0.53) and Typomind (0.31). This demonstrates that Typomind's name-only analysis misses over two-thirds of real-world attacks, while ConfuGuard's heuristic-based filter still misses almost half. AgentGuard's ability to avoid these false negatives is critical in security contexts where missing a threat carries severe consequences.

On NeupaneDB, which contains more clear-cut, publicly documented attacks, all methods perform better, but AgentGuard maintains its advantage with a weighted F1 of 0.93 versus 0.76 (ConfuGuard) and 0.69 (Typomind). The smaller performance gap on this dataset suggests that AgentGuard's advantages are most pronounced when handling subtle, analyst-triaged real-world cases, precisely the challenging scenarios where automated detection provides the most value.

\textit{\textbf{Understanding the Performance Gains:}} AgentGuard's superior performance stems from key design choices that directly address the fundamental limitations of prior work: 

\textit{a) Proactive and Hybrid Target Discovery:} Unlike the baselines' single-signal retrieval, the Threat Analyst Agent's hybrid strategy is demonstrably superior at finding the correct target (\textbf{G1}), superior to the name-only analysis of Typomind, which fails to detect a high volume of attacks (e.g., 0.31 Confusion Recall on ConfuDB).

\textit{b) Fused Feature Model:}
Our model overcomes Typomind's limitations by incorporating critical metadata and code context. It simultaneously surpasses ConfuGuard by replacing its limited heuristic filter with a full-fledged ML model. This fused model integrates all signals including temporal signal features (TS) for robustness and package content analysis to correctly classify benign packages that are misidentified by metadata-only approaches.

\textit{c) Data-Driven Thresholding:} The ML classifier is paired with an empirically optimized threshold (0.42, from F1-maximization). This allows AgentGuard to be operationally tuned to favor recall, unlike baselines whose fixed rules or heuristics lead to poor and inflexible recall (as seen with baselines' 0.53 and 0.31 Confusion Recall on ConfuDB).

\textit{\textbf{Answer to RQ2:}} \approach{} significantly outperforms baseline methods with by up to 17\%-43\% F1 score improvement and 11\%-35\% FPR reduction.


\subsection{RQ3 - Hybrid Search Target Discovery Accuracy}
To analyze the performance of AgentGuard's hybrid search strategy which fuses semantic and syntactic search against its constituent parts. We aim to demonstrate that this fusion is necessary to achieve high recall across diverse confusion types (lexical, syntactic, and semantic). We use the ground-truth pairs from the non-benign subsets of the ConfuDB and NeupaneDB. We implemented three search strategy in \fref{fig:tdr}: \textbf{Semantic} (only Cosine Similarity), \textbf{Syntactic} (only Trigram Similarity) and \textbf{Hybrid} (complete hybrid search, AgentGuard applied). As the primary metric, we use Target Discovery Rate (TDR@k) measures the percentage of malicious packages for which the algorithm successfully retrieved the correct legitimate target within the top $k$ results.
\textit{TDR@1} measures final target discovery precision. \textit{TDR@3} measures effectiveness of the legitimate-package filtering and ranking. \textit{TDR@30} measures total recall, whether the target found at all in the initial candidate pool.

\begin{figure}[tbp]
    \centering
    \includegraphics[width=\columnwidth]{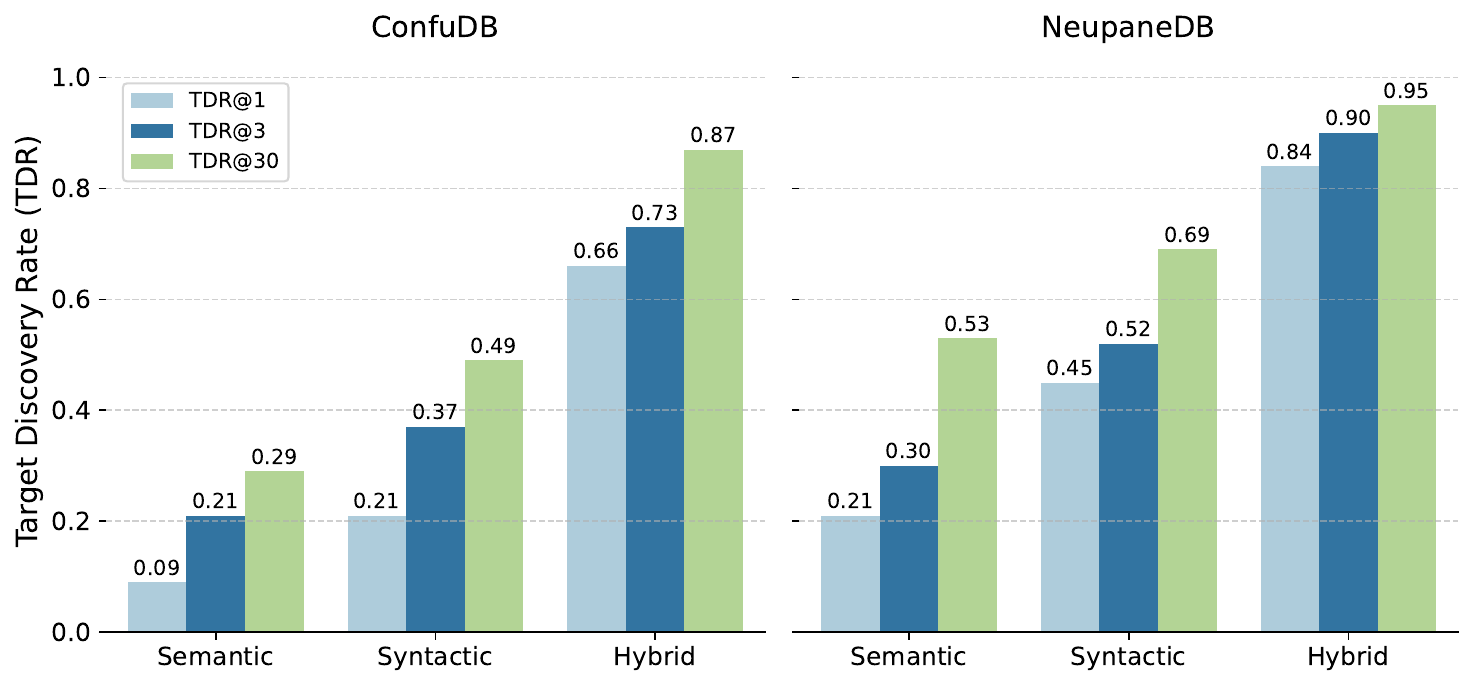}
    \caption{Target Discovery Rate (TDR@k) comparison of the three search strategies. Hybrid Search significantly outperforms both single-signal models across all metrics.}
    \label{fig:tdr}
\end{figure}

\textit{\textbf{Limitations of Single-Signal Strategies:}} Both Semantic-Only and Syntactic-Only models exhibit critical, complementary failures. Syntactic-Only struggles with total recall (TDR@30), scoring poorly on both datasets (0.49 and 0.69). This is because it is structurally blind to semantic attacks, which share no character-level similarity. Semantic-Only performs poorly on precision (TDR@1), scoring lowest on both datasets (0.09 and 0.21). It successfully recalls semantic neighbors but often fails to rank the most dangerous, lexically-similar typos (e.g., requests vs reqeusts) as the top result, prioritizing other, less-threatening synonyms instead.

\begin{table*}[t!]
\centering
\small
\caption{Ablation experiment results comparing six model configurations on the ConfuDB and NeupaneDB datasets. Performance is evaluated on Clean data, and robustness is measured against adversarial (Adv.) attacks targeting MQ features. }
\label{tab:ablation_study}

\renewcommand{\arraystretch}{1.2} 

\begin{tabularx}{\textwidth}{@{} >{\raggedright\arraybackslash}p{1.8cm} *{10}{>{\centering\arraybackslash}X} @{}}
\toprule

\multirow{2}{*}{\textbf{Model}} 
& \multicolumn{5}{c}{\textbf{ConfuDB}} & \multicolumn{5}{c}{\textbf{NeupaneDB}} \\ \cmidrule(lr){2-6} \cmidrule(lr){7-11}
& AUC$_{(clean)}$& F1$_{(clean)}$ & Recall$_{(clean)}$ & Recall$_{(adv.)}$ & $\Delta \text{Recall}$
& AUC$_{(clean)}$& F1$_{(clean)}$ & Recall$_{(clean)}$ & Recall$_{(adv.)}$ & $\Delta \text{Recall}$ \\
\midrule
SS-Only & 0.58 & 0.58 & 0.56 & 0.56 & 0.00 & 0.80 & 0.75 & 0.74 & 0.74 & 0.00 \\

SS+MQ & 0.78 & 0.74 & 0.60 & 0.41 & 0.19
& 0.95 & 0.92 & 0.91 & 0.11 & 0.80 \\
SS+CD+TS & 0.77 & 0.74 & 0.53 & 0.53 & 0.00
& 0.95 & 0.91 & 0.90 & 0.90 & 0.00 \\
SS+MQ+CD+TS & 0.82 & 0.77 & 0.60 & 0.45 & 0.16
& 0.96 & 0.93 & 0.91 & 0.73 & 0.19 \\
SS+CD+TS+PC & 0.80 & 0.78 & 0.61 & 0.61 & 0.00
& 0.96 & 0.92 & 0.92 & 0.92 & 0.00 \\

\rowcolor{lightgray} 
\textbf{All-Features} & \textbf{0.83} & \textbf{0.80} & \textbf{0.64} & \textbf{0.43} & \textbf{0.20}
& \textbf{0.96} & \textbf{0.93} & \textbf{0.92} & \textbf{0.85} & \textbf{0.07} \\

\bottomrule
\end{tabularx}
\end{table*}

\textit{\textbf{Hybrid Strategy Excels in Accuracy:}} The AgentGuard-Hybrid model dominates across all metrics, demonstrating its superiority. While its total recall (TDR@30) of 87\%-95\% proves it successfully captures targets from both semantic and syntactic attack vectors (solving the recall limitations of Syntactic-Only), the most critical metric is TDR@1.

In our multi-agent systems, the Threat Analyst Agent is responsible for identifying the single most likely target before passing it to the Confusion Checker Agent for expensive package content analysis. Therefore, TDR@1 directly measures the success rate of this handoff.

The AgentGuard-Hybrid model achieves a TDR@1 of 66\%-84\%, which is significantly higher than both Semantic-Only (9\%-21\%) and Syntactic-Only (21\%-45\%). This demonstrates that our hybrid ranking score is highly effective at balancing semantic and syntactic signals, ensuring that the correct legitimate target is prioritized as the Top-1 candidate in the vast majority of cases, which is essential for the subsequent content analysis stage to function correctly (\textbf{G1}).

\textit{\textbf{Answer to RQ3:}} AgentGuard's hybrid search strategy is essential, as it is the only approach that achieves both the high recall and high precision necessary to power the end-to-end detection pipeline.

\subsection{RQ4 - Feature Contribution and Robustness}
To quantify the contribution of each feature category to overall performance, and to assess their respective impact on model resilience against adversarial manipulation, we expanded our controlled ablation experiments to compare six distinct model configurations: \textbf{SS-Only}, \textbf{SS+MQ} (SS and MQ), \textbf{SS+CD+TS} (SS, CD and TS, excluding MQ), \textbf{SS+MQ+CD+TS} (SS, MQ, CD and TS), \textbf{SS+CD+TS+PC} (SS, CD, TS and PC) and \textbf{All-Features} (AgentGuard). Each model was evaluated on two test conditions using the NeupaneDB and ConfuDB feature sets: \textit{Clean}—The original, unmodified test data. \textit{Adversarial (Adv.)}—A simulated attack where all forgeable Metadata Quality (MQ) feature values in \textit{Confusion} samples are flipped to appear benign (e.g., repo\_valid=1), while \textit{Benign} samples remain unchanged.
The key metric is the recall drop, $\Delta \text{Recall} = Recall_{\text{Clean}} - Recall_{\text{Adv.}}$ for the \textit{Confusion} class, quantifying performance degradation under attack. \tref{tab:ablation_study} presents the complete results.

\textit{\textbf{Feature Group Contribution:}} The baseline SS-Only establishes low-performance benchmarks (F1 = 0.58 ConfuDB, 0.75 NeupaneDB), confirming syntactic similarity as an insufficient signal. Performance significantly increases by integrating metadata: adding either MQ features (SS+MQ, F1=0.74/0.92) or robust CD/TS features (SS+CD+TS, F1=0.74/0.91) yields substantial gains on both datasets. This underscores the high predictive value of metadata signals. Crucially, Package Content (PC) features provide a complementary and distinct performance lift; augmenting SS+CD+TS with PC features (creating SS+CD+TS+PC) boosts the F1-score from 0.74 to 0.78 on ConfuDB and from 0.91 to 0.92 on NeupaneDB. Finally, the AgentGuard (All-Features, including all features) achieves the highest baseline performance, yielding an F1-score of 0.93 on NeupaneDB and 0.80 on ConfuDB, demonstrating that the fusion of all five feature groups captures the most comprehensive view of package risk.

\textit{\textbf{Robustness Against Adversarial Manipulation:}} The adversarial evaluation reveals the critical role of feature design in building a resilient classifier (\textbf{G3}).

\textit{a) Metadata-Only Vulnerability:} The SS+MQ (SS and MQ), which relies on easily-forged features, suffers a catastrophic performance collapse on NeupaneDB, where its $Recall_{\text{Clean}}$ of 0.91 plummets to 0.11 (a 0.80 Recall Drop). This vulnerability, while less pronounced, is also evident on ConfuDB, where its recall drops from 0.60 to 0.41 (a 0.19 drop). This confirms that detectors relying heavily on this type of metadata are fragile and can be trivially bypassed.

\textit{b) The Content-Agnostic Trade-off:} The SS+CD+TS (SS, CD and TS), which replaces forgeable MQ features with robust temporal signals, is perfectly immune to this specific metadata attack on both datasets. However, this immunity comes at a cost to baseline performance. Its F1-score (0.74 on ConfuDB / 0.91 on NeupaneDB) is consistently lower than the metadata-complete SS+MQ+CD+TS (0.77 on ConfuDB / 0.93 on NeupaneDB). This demonstrates a fundamental trade-off for metadata-only detectors: one must choose between maximum performance (using MQ) or maximum robustness (omitting MQ).

\textit{c) Breaking the Trade-off with Package Content:} For the SS+MQ+CD+TS (only metadata), although its TS provide a \enquote{defensive hedge}, this content-agnostic model remains vulnerable to the metadata forgery attack, suffering significant Recall Drops of 0.16 (ConfuDB) and 0.19 (NeupaneDB). The All-Features (AgentGuard) fuses the same 14 metadata features with our 4 PC features. The impact of this fusion is most dramatic on NeupaneDB, where resilience improves significantly: the Recall Drop plummets by nearly 3x, from 0.19 (SS+MQ+CD+TS) to just 0.07 (All-Features). On ConfuDB, the drop is 0.20, comparable to SS+MQ+CD+TS's 0.16. The above results demonstrate that the PC features act as a powerful, independent verification layer. When the MQ features are forged to appear benign, All-Features can rely on the strong, conflicting signals from the package content analysis. This allows it to \enquote{override} the forged metadata and maintain a high recall of 0.85 on NeupaneDB (a significant improvement over SS+MQ+CD+TS's 0.73). This content-aware design successfully breaks the metadata-only trade-off, achieving both the highest baseline performance and the best-in-class robustness.

\begin{figure}[tbp]
    \centering
    \includegraphics[width=\columnwidth]{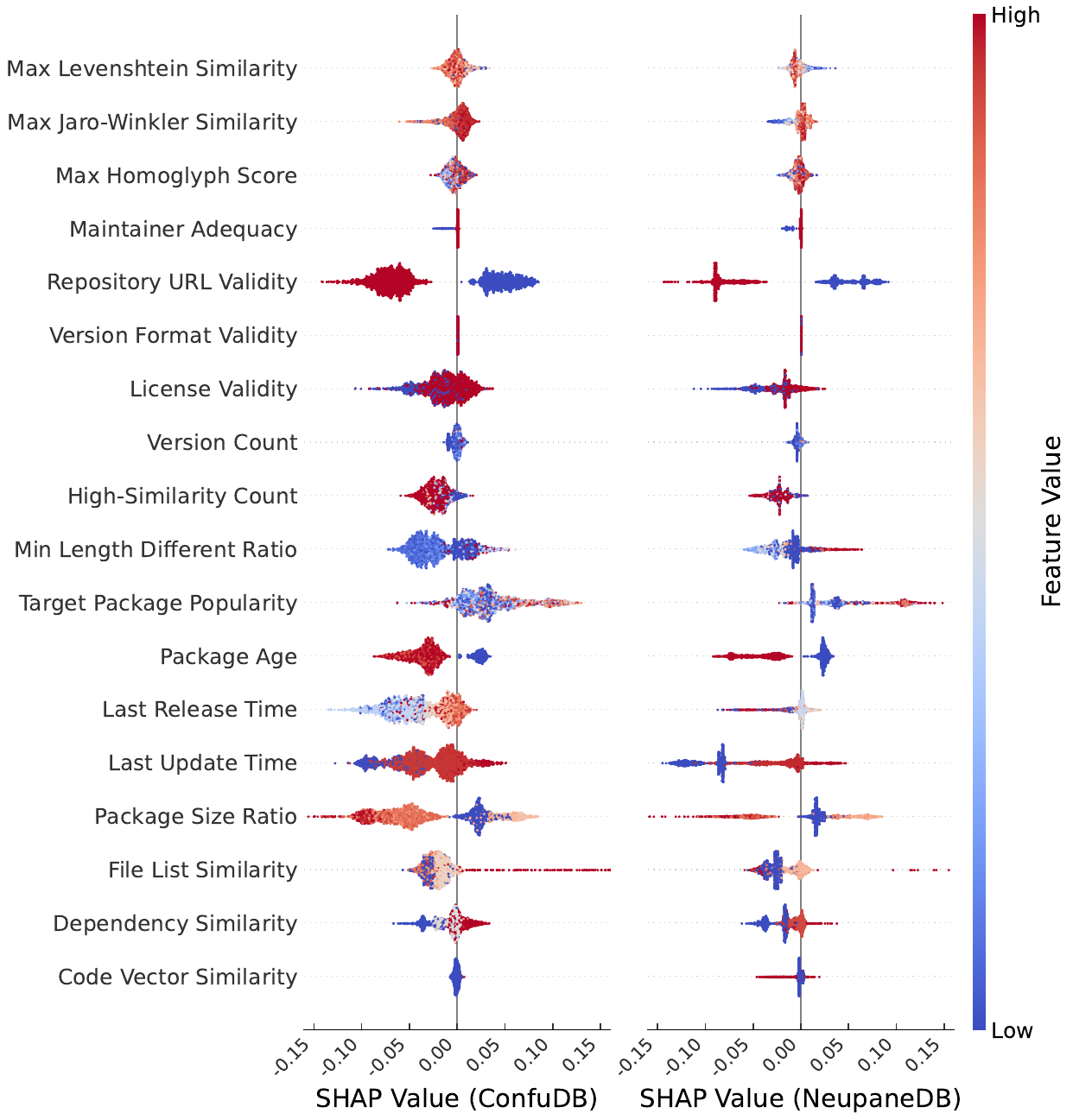}
    \caption{SHAP dot plot showing the contribution of all 18 features (ordered by SS, MQ, CD, TS, and PC groups) to the model's prediction on ConfuDB and NeupaneDB.}
    \label{fig:sharp}
\end{figure}

\textit{\textbf{Model Interpretability:}} To understand why the fused model (All-Features) is both high-performing and robust, we analyzed its internal logic using SHAP Analysis~\cite{shap}. \fref{fig:sharp} plots the SHAP values for all 18 features across both datasets. The X-axis represents the feature's impact on the prediction (positive values push towards \textit{Confusion}), and the color represents the feature's value.
The plot reveals two distinct types of features:

\textit{a) Risk Indicators:} For features like Max Jaro-Winkler Similarity, Maintainer Adequacy, License Validity and Dependency Similarity high values (red dots) correspond to positive SHAP values. This demonstrates that the model has learned to associate these signals with increased risk. For example, high Dependency Similarity (red dots), rather than indicating trust, is pushed towards \textit{Confusion} (positive SHAP), suggesting the model identifies that malicious mimics often copy dependency files directly. Similarly, high values for Maintainer Adequacy and License Validity (red dots) pushing towards \textit{Confusion} reveals the model's reliance on other features to override these easily forgeable metadata signals when they appear benign in a malicious package.

\textit{b) Trust Indicators:} For features like Repository URL Validity, High-Similarity Count, Package Age and Package Size Ratio, the pattern is reversed. High values (red dots) correspond to negative SHAP values. This confirms the model's core logic: high (red) feature values such as a valid repository url, a high package age, or a high package size ratio are strong signals of legitimacy that push the model's prediction towards \textit{Benign} (negative SHAP). Conversely, low (blue) values for these features (e.g., a missing repository, a new package, or highly dissimilar code size) are treated as significant risk factors, pushing the prediction towards \textit{Confusion}.

When MQ features (like Maintainer Adequacy) are adversarially flipped from Low (blue) to High (red), the attacker successfully generates a slight push towards \textit{Confusion}. However, this manipulation is rendered ineffective because the TS and PC features (like Package Size Ratio and Package Age) remain Low (blue), generating a much stronger, opposing push towards \textit{Confusion} (positive SHAP). This multi-dimensional features allows All-Features to override the forged metadata, demonstrating that the TS and PC features act as a powerful, independent verification layer.

\textit{\textbf{Answer to RQ4:}} All feature categories contribute to detection performance, highlighting that a fused, multi-dimensional design is the most effective and robust architecture for \approach{}.

\section{Discussion}
\label{sec:discussion}

Our empirical evaluation validates AgentGuard as a proactive and robust framework. We now synthesize these findings to discuss the implications of our design choices.

\subsection{Breaking the Robustness Trade-off via Content Fusion}
Our ablation study reveals that fused metadata-code analysis solves a fundamental stalemate faced by prior content-agnostic detectors: the choice between performance (trusting forgeable MQ features) and robustness (discarding them).
Our results confirm that while metadata-only models suffer catastrophic collapse under evasion (e.g., SS+MQ dropped 0.80 in recall), the integration of PC features decisively breaks this trade-off. Adding PC signals reduced the adversarial recall drop by nearly $3\times$ (from 0.19 to 0.07) compared to the robust metadata baseline. 
Crucially, as revealed by our SHAP analysis, package content acts as a non-forgeable verification layer. When metadata is forged, conflicting content signals effectively override the deception, allowing AgentGuard to achieve best-in-class robustness without sacrificing baseline performance.

\subsection{The Necessity of Hybrid Search for Proactive Detection}
Proactive detection is viable only if the system autonomously identifies the correct victim. We demonstrate that single-signal strategies are insufficient: Semantic-Only lacks precision (low TDR@1), while Syntactic-Only lacks recall (low TDR@30).
AgentGuard's hybrid strategy is the only approach to achieve high discovery rates across all metrics, crucially reaching a TDR@1 of 73-80\%. This high precision is architecturally critical: it ensures that the computationally expensive Package Content Analysis is executed solely on the most probable target pair, validating the efficiency and feasibility of our Threat Analyst Agent design.

\subsection{Rationale for Multi-Agent Architecture}
\label{subsec:multi_agent_rationale}
The adoption of a multi-agent architecture, rather than a monolithic pipeline, is necessitated by two key system characteristics:

\textit{Cognitive Specialization via Tool Augmentation:} Unlike static software modules, our agents act as specialized decision-makers. The \textit{Threat Analyst} leverages LLM-driven reasoning to autonomously pinpoint the single most probable victim from the most similar targets set, while the \textit{Confusion Checker} employs discriminative ML for precise verification. This architecture allows us to seamlessly bridge probabilistic reasoning with deterministic evaluation within a unified workflow.

\textit{Adaptive Orchestration and Resilience:} The agent-based design enables dynamic control flow beyond rigid pipelines. The \textit{Orchestrator} autonomously manages context switching across 34 ecosystems and executes resilient fallback strategies. This capability allows the system to adapt to new confusion patterns via prompt engineering rather than code refactoring, ensuring superior extensibility.

\subsection{Limitations and Future Work}
\label{sec:limitations}

AgentGuard faces two primary limitations: first, while our temporal signals resist rapid metadata forgery, they remain vulnerable to sophisticated sustained strategies like account nurturing, which future work aims to mitigate via author-level reputation signals. Second, our content analysis targets impersonation rather than stealthy injection; consequently, attacks embedding small payloads into legitimate code-based (e.g., \texttt{event-stream}) may evade detection, necessitating the future integration of fine-grained static analysis or dependency auditing.

\section{Conclusion}
\label{sec:conclusion}
This paper presented \approach{}, a novel multi-agent framework for proactive package confusion detection. By introducing a pioneering single-input active detection paradigm, \approach{} overcomes the critical limitations of prior work, which were restricted to single-signal retrieval strategies (relying solely on lexical or semantic metrics) and lacked deep package content analysis. Through the synergy of hybrid search target discovery and content-aware multi-dimensional evaluation, our framework effectively identifies complex confusion attacks, significantly outperforming state-of-the-art baselines. Ultimately, \approach{} provides a practical and adversarially robust defense mechanism, crucial for safeguarding the modern software supply chain against increasingly sophisticated threats.

\bibliographystyle{unsrt}
\bibliography{references.bib}

\end{document}